\documentclass[12pt,preprint2]{emulateapj}

\slugcomment{submitted to ApJ Letter}

\shorttitle{Relativistic Magnetic Reconnection}
\shortauthors{Takahashi et al.}

\begin{document}

\title{Scaling Law of Relativistic Sweet--Parker Type Magnetic Reconnection}

\author{Hiroyuki R. Takahashi \altaffilmark{1}}
\affil{Center for Computational Astrophysics, National
  Astronomical Observatory of Japan, 2-21-1, Osawa, Mitaka, Tokyo
  181-8588, Japan}
\email{takahashi@cfca.jp}
\author{Takahiro Kudoh \altaffilmark{2}}
\affil{Division of Theoretical Astronomy, National Astronomical
Observatory of Japan, 2-21-1, Osawa, Mitaka, Tokyo 181-8588, Japan}
\author{Youhei Masada\altaffilmark{3}}
\affil{Department of Computational Science, Kobe University,
1-1 Rokkodai, Nada, Kobe 657-8501, Japan}
\and
\author{Jin Matsumoto \altaffilmark{4}}
\affil{Kwasan and Hida Observatories, Graduate School of Science, Kyoto
University, Kyoto 607-8471, Japan}

\begin{abstract}
Relativistic Sweet--Parker type magnetic reconnection is investigated by 
relativistic resistive magnetohydrodynamic (RRMHD) simulations. 
As an initial 
setting, we assume anti-parallel magnetic fields and a spatially uniform
resistivity.
A perturbation imposed on the magnetic fields
triggers magnetic reconnection around a current sheet, and the plasma
inflows into the reconnection region. 
The inflows are then heated due to ohmic dissipation in the
 diffusion region, 
and finally become relativistically hot outflows. 
The outflows are not accelerated to ultra-relativistic speeds (i.e., Lorentz factor
 $\simeq 1$),
even when the magnetic energy dominates the thermal and rest mass energies in the inflow region. 
Most of the magnetic energy in the inflow region is converted into the thermal energy of the outflow
during the reconnection process.
The energy conversion from magnetic to thermal energy in the diffusion region
results in an increase in the plasma inertia.
This prevents the outflows from being accelerated to ultra-relativistic speeds.
We find that the reconnection rate $\mathcal{R}$ obeys the scaling relation 
$\mathcal{R} \simeq S^{-0.5}$, where $S$ is the Lundquist number.
This feature is the same as that of non-relativistic reconnection.
Our results are consistent with the theoretical predictions of \cite{2005MNRAS.358..113L} for 
Sweet--Parker type magnetic reconnection. 
\end{abstract}

\section{Introduction}\label{intro}
Magnetic reconnection is one of the 
most important subjects in the studies of space, laboratory, and
astrophysical plasmas 
\citep{1986PhFl...29.1520B, 2000mare.book.....P}. 
In particular, it plays
an essential role in the understanding of energy conversion 
processes in high energy plasmas that characterize astrophysical compact objects, such as neutron stars 
\citep{1984ApJ...283..710K, 2001ApJ...547..437L, 2003MNRAS.339..765L}, soft gamma-ray repeaters 
\citep{2006MNRAS.367.1594L, 2010MNRAS.407.1926G, 2010PASJ...62.1093M}, active galactic nuclei
\citep{1998MNRAS.299L..15D}, and gamma-ray bursts
\citep{2002A&A...387..714D, 2010arXiv1011.1904M, 2011ApJ...726...90Z}.

While issues concerning the physical mechanisms and properties of
magnetic reconnection
remain unsettled in the relativistic regime,
a few theoretical studies of relativistic effects 
have been made \citep{1994PhRvL..72..494B,
2003ApJ...589..893L, 2005MNRAS.358..113L, 2007AdSpR..40.1538T,
2010MNRAS.403..335T}.
 \cite{2003ApJ...589..893L} studied 
relativistic Sweet--Parker type reconnection
within the framework of magnetohydrodynamics (MHD), finding that the
reconnection-driven outflow can have an ultra-relativistic speed (Lorentz
factor $\gamma \gg 1$) when the magnetic energy is preferentially
converted to kinetic energy. They concluded that the reconnection rate would be
enhanced in the relativistic regime due to Lorentz contraction. 
In contrast, \cite{2005MNRAS.358..113L} concluded that the outflow
cannot be accelerated to a relativistic speed ($\gamma \simeq 1$)
because the magnetic
energy should be converted into thermal energy to maintain
the pressure balance across the current sheet. The effect of the Lorentz
contraction is then negligible and the reconnection rate would
not be enhanced. 
\begin{deluxetable*}{lccccccccccc}[!h]
\tabletypesize{\scriptsize}
\tablecaption{List of Simulation Runs}
\tablewidth{0pt}
\tablehead{
\colhead{model} & \colhead{$\beta_0$} & \colhead{$\sigma_0$} &
 \colhead{$u_{A,0}$} & \colhead{$R_M$} & \colhead{$\gamma_\mathrm{max}$}
 & \colhead{$\tilde{M}_{A,\mathrm{max}}$} & \colhead{$t_{20}$} &
\colhead{$(1+h+\sigma)|_\mathrm{in}$} & \colhead{$\sigma_\mathrm{out}$} &
\colhead{$h_\mathrm{out}$} & \colhead{$\gamma_\mathrm{out}$}
}
\startdata
  B1R2  & 0.1 & 20  & 2.00 & 200 & 5.0 & 2.45
&191.4  & 22.7 & $1.86 \times 10^{-2} $& 33.7 &1.40\\
  B2R2  & 0.2 & 10  & 1.41 & 200 & 3.0 & 2.00
& 195.1 & 13.8 & $6.58 \times 10^{-3} $& 18.7  &1.29\\
  B4R2  & 0.4 & 5   & 1.00 & 200 & 2.0 & 1.73
& 206.5 & 9.36 & $2.22 \times 10^{-3} $& 11.1  &1.17\\
  B8R2  & 0.8 & 2.5 & 0.71 & 200 & 1.5 & 1.58
& 233.3 & 7.15 & $8.27 \times 10^{-4} $& 7.48 &1.08\\
  B2R1  & 0.2 & 10  & 1.41 & 100 & 3.0 & 2.00
& 290.7 & 13.4 & $9.56 \times 10^{-3} $& 19.5 &1.20\\
  B2R5  & 0.2 & 10  & 1.41 & 500 & 3.0 & 2.00
& 157.6 & 14.1 & $2.21 \times 10^{-3} $& 13.9 &1.28\\
  B2R10 & 0.2 & 10  & 1.41 &1000 & 3.0 & 2.00
& 151.9 & 14.4 & $1.00\times 10^{-3} $& 9.91 &1.22
\enddata
\tablecomments{Columns: (6) possible maximum outflow Lorenz factor
 evaluated from equation (\ref{eq:gmax}); (7) $\tilde{M}_{A,\mathrm{max}}\equiv
 u_\mathrm{max}/u_{A,0}$; (8) time at which the current sheet length
 reaches $Y=20\equiv L_{20}$; (9) the total specific enthalpy evaluated at
 $(X,Y)=(5,0)$ when $t=t_{20}$; (10)-(12) the magnetization parameter, the
 specific enthalpy, and the Lorentz factor evaluated at
 $(X,Y)=(0,L_{20})$ when $t=t_{20}$.}
\label{tab:param}.
\end{deluxetable*}

Recently, further numerical studies have been conducted on
relativistic magnetic reconnection. Particle-in-cell simulations are
mainly employed
to ascertain the reconnection mechanism in the collisionless regime without introducing a phenomenological 
parameter, i.e., electric resistivity  
\citep{2001ApJ...562L..63Z, 2004ApJ...605L...9J, 2007ApJ...670..702Z, 2008ApJ...684.1477Z, 2007PhPl...14e6503B, 
2008PhPl...15b2101Z}. 
\cite{2006ApJ...647L.123W} studied
Petschek type relativistic magnetic reconnection for the first time 
in a spatially localized resistivity model using relativistic resistive
magnetohydrodynamic (RRMHD) simulations. 
Relativistic two-fluid MHD simulations were performed by \cite{2009ApJ...705..907Z, 2009ApJ...696.1385Z}.
\cite{2010ApJ...716L.214Z} numerically examined relativistic
magnetic reconnection using various resistivity models and obtained 
Petschek and Sweet--Parker type magnetic reconnection in relativistic
plasmas. 

While these previous studies have revealed much about Petschek type
reconnection in the regime of low magnetic Reynolds number, $R_M \lesssim 160$,
the physics of Sweet--Parker type magnetic reconnection and the
dependence of the reconnection rate on the magnetic Reynolds number
remain unsolved. 
In this Letter, we focus on Sweet--Parker type magnetic
reconnection and investigate the basic properties through RRMHD simulations.
This is the first systematic study of relativistic Sweet--Parker type
magnetic reconnection for a broad range of magnetic Reynolds number $R_M\leq
10^3$.

\section{Numerical Model}\label{model}
We numerically solve a set of RRMHD equations in two-dimensional Cartesian coordinates ($X$,$Y$) 
with the simple form of Ohm's law,
\begin{equation}
 \mbox{\boldmath$j$} = q\mbox{\boldmath$v$}+\eta^{-1}
  \gamma[\mbox{\boldmath$E$}+\mbox{\boldmath$v$}\times
  \mbox{\boldmath$B$}-(\mbox{\boldmath$E$}\cdot
  \mbox{\boldmath$v$})\mbox{\boldmath$v$}],
  \label{eq:ohm}
\end{equation}
where $q$, $\mbox{\boldmath$j$}$, $\eta$, $\mbox{\boldmath$v$}$,
$\gamma$, $\mbox{\boldmath$E$}$, $\mbox{\boldmath$B$}$ are the charge 
density, electric current, electric resistivity, three velocity, Lorentz
factor, electric field, and magnetic field, respectively.
We set the light speed, Boltzmann constant, and average particle
mass as unity throughout this paper. 

The relativistic Harris sheet is adopted as an initial setting \citep{2003ApJ...591..366K}. 
The spatial distribution of the magnetic field is then given by 
$B_x(X,Y)=B_z(X,Y)=0$ and $B_y(X,Y)=B_0\tanh(2X/\lambda)$, where $B_0$ 
is the field strength of the sheath plasma $(X\rightarrow \pm \infty)$ and
$\lambda$ is the initial thickness of the current sheet. 
The gas pressure and density profiles are determined by the local pressure balance, 
i.e. $p=p_0+[B_0^2-B_y^2(X,Y)]/(8\pi)$ and $\rho=p/T_0$,
where $p_0$ is the initial gas pressure of the sheath plasma 
and $T_0$ is the initial plasma temperature, which is assumed to be
constant throughout the region. 
We set $p_0=T_0=1$
and fix the specific heat ratio $\Gamma$ 
as $\Gamma=4/3$ throughout this paper. 
The magnetic reconnection is triggered around the origin by a 
perturbation of the magnetic field described in vector potential 
form as $\delta A_z=- \delta B_0\lambda \exp[-(X^2+Y^2)/\lambda^2]$, 
where $\delta B_0=0.03B_0$ is the amplitude of the perturbation.

The plasma $\beta$ in the sheath is defined as $\beta_0 \equiv 8\pi p_0/B_0^2$. 
The magnetization parameter corresponding to each model is $\sigma_0 =20$, 10, 5, 2.5 , 
where $\sigma\equiv B^2/(4\pi\rho \gamma^2)$ and a subscript \lq \lq 0''
denotes a quantity of the sheath plasma.
The corresponding Alfv\'en four speed is $u_{A,0}\equiv
v_{A,0}/\sqrt{1-v_{A,0}^2}=2.0$, 1.41, 1.0, 0.71,
where $v_{A}$ is the Alfv\'en speed.
In our models, the electric resistivity is assumed to be uniform.
We vary the resistivity $\eta$ for magnetic Reynolds number 
$R_M\equiv 4\pi \lambda / \eta=100$, 200, 500, 1000 . 
We note that $R_M$ defined by the full thickness of the initial current
sheet is twice as large as that used in
the previous study by \cite{2010ApJ...716L.214Z} who defined
$R_M$ by the half thickness of the current sheet. 
The parameters adopted in each model are summarized in Table~\ref{tab:param}.

The two-dimensional calculation is performed in the $X$--$Y$ plane with a
volume bounded by $X=[0, 50]$ and $Y=[0, 75]$. 
The length and time are normalized by the initial thickness of the
current sheet $\lambda $ and its Alfv\'en crossing time $\tau_A\equiv
\lambda/v_{A,0}$. 
We use a non-uniform
grid in the $X$-direction and a uniform grid in the $Y$-direction 
 of $790\times 3000$ zones. The minimum grid size in each direction is $\Delta x = 5\times 10^{-3}$ and
$\Delta y=2.5\times 10^{-2}$. 
A symmetric boundary condition is applied at $X=0$ and $Y=0$. A free 
boundary condition is imposed at $X=50$ and $Y=75$.
We calculate numerical fluxes using the HLL method \citep{1983siamRev...25..35..61,
2007MNRAS.382..995K}
with an operator-splitting method. 
We use the implicit scheme to solve Ampere's equation 
in order to maintain numerical stability when the electric resistivity is small \citep{2009MNRAS.394.1727P}.  
The implicit scheme enables us to study magnetic reconnection with 
larger magnetic Reynolds number than previous studies.
\begin{figure*}
   \center
   \includegraphics[height=7.0cm]{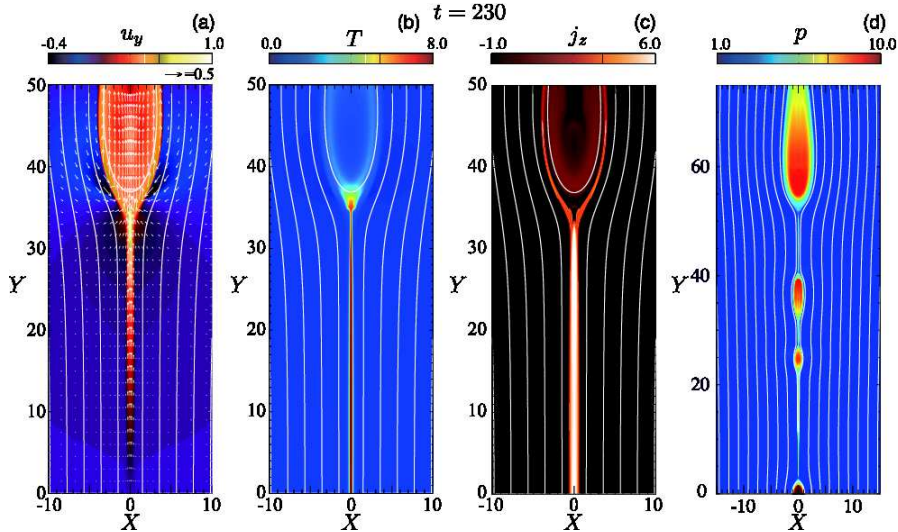}
 \caption{Snapshots at $t=230$ of models B2R2 (Figs. 1a, 1b, and 1c) and
 B2R10 (Fig. 1d). 
The colors show (a) the outflow component of the four velocity ($u_y$),
 (b) the plasma temperature ($T = p/\rho$), (c) the
electric current perpendicular to the $X$--$Y$ plane ($j_z$), and (d) the
 gas pressure ($p$). The solid lines show the magnetic field lines. 
The arrows in Fig. 1a represent the velocity fields.
 }
 \label{fig:fig1}
\end{figure*}

Before presenting the numerical results, 
we estimate the speed of outflow driven by magnetic reconnection.
Consider steady Sweet--Parker type magnetic reconnection 
in a diffusion region of length $2\delta $ in the inflow direction and $2L$ 
in the outflow direction. 
Mass and energy conservation between the inflow and outflow give 
\begin{equation}
 \rho_{\mathrm{in}} u_{\mathrm{in}} 2L 
= \rho_\mathrm{out} u_\mathrm{out} 2\delta,
\end{equation}
\begin{eqnarray}
 &&\left[\left(\rho_{\mathrm{in}} + \frac{\Gamma}{\Gamma-1} p_{\mathrm{in}}\right)
  \gamma_{\mathrm{in}} u_{\mathrm{in}} 
  + \frac{v_{\mathrm{in}} B_{\mathrm{in}}^2}{4\pi} \right]2L\nonumber \\
&&=\left[\left(\rho_\mathrm{out} + \frac{\Gamma}{\Gamma-1} p_\mathrm{out}\right)
  \gamma_\mathrm{out} u_\mathrm{out} 
  + \frac{v_\mathrm{out} B_\mathrm{out}^2}{4\pi} \right]2\delta,
\end{eqnarray}
where $\mbox{\boldmath$u$} = \gamma \mbox{\boldmath$v$}$ is the four velocity. 
The subscripts \lq \lq in''
and \lq \lq out'' indicate physical quantities 
in the inflow and outflow regions.
Note that the ideal MHD
($\mbox{\boldmath$E$} = -\mbox{\boldmath$v$} \times \mbox{\boldmath$B$}$)
can describe both inflow and outflow plasmas because these flows are 
outside the diffusion region.
Combining equations~(2) and (3), we obtain
\begin{equation}
 \left(1 + \sigma_{\mathrm{in}} + h_{\mathrm{in}}\right)\gamma_{\mathrm{in}} 
= \left(1 + \sigma_\mathrm{out} + h_\mathrm{out}\right)\gamma_\mathrm{out},
 \label{eq:bernoulli}
\end{equation}
where $h=\Gamma p/[(\Gamma-1)\rho]$ is the specific 
enthalpy. This is similar to the Bernoulli equation 
that describes the conservation of the total enthalpy flux
between the inflow and outflow. 
Assuming $\sigma_\mathrm{out} \ll 1$, which is
reasonable for the case of
anti-parallel ($B_z=0$) magnetic reconnection, we obtain
\begin{equation}
 \gamma_\mathrm{out} = \frac{1 + h_\mathrm{in} + \sigma_\mathrm{in}}{1 +
  h_\mathrm{out}}\gamma_\mathrm{in}.\label{eq:gout}
\end{equation}
If the magnetic energy is converted preferentially
into kinetic energy ($h_\mathrm{in}= h_\mathrm{out}$), as was ideally
assumed in
\cite{2003ApJ...589..893L}, we obtain an upper limit of the outflow Lorentz
factor $\gamma_\mathrm{max}$,
\begin{equation}
 \gamma_\mathrm{max} = \frac{1 + \sigma_{\mathrm{in}} + h}{1 +
  h}\gamma_{\mathrm{in}} \;.
  \label{eq:gmax}
\end{equation}
$\gamma_\mathrm{max}$ listed in
Table~\ref{tab:param} is evaluated 
from an initial $\sigma_\mathrm{in}$ and $h$ (we take $\gamma_\mathrm{in}=1$ in all models for
numerical evaluation). 
Equation (\ref{eq:gmax}) reduces to
$\gamma_\mathrm{max}=(1+\sigma_{\mathrm{in}})\gamma_{\mathrm{in}}$
when the thermal energy is negligible (see also equation 32 in
\citealt{2003ApJ...589..893L}). The outflow becomes super-Alfv\'enic in
this case.

We stress that $\gamma_\mathrm{max}$ is the upper limit of the
outflow Lorentz factor under the condition $h_\mathrm{in}=h_\mathrm{out}$.
Since a part of the magnetic energy should be spent for plasma heating
by ohmic dissipation, the enthalpy of the outflow is expected to 
increase and its Lorentz factor would be less than
$\gamma_\mathrm{max}$ except in the ideal situation.

\section{Results} \label{sec:result}
Figure~\ref{fig:fig1} shows snapshots at $t=230$ of model B2R2
(Figs. 1a, 1b, and 1c) and B2R10 (Fig. 1d). We focus here on Fig. 1a--1c
(model B2R2), and will refer to Fig. 1d in \S~\ref{sec:conclusion}.
The colors show (a) the outflow component of
the four velocity ($u_y$), (b) plasma temperature ($T=p/\rho$), 
(c) electric current perpendicular to the $X$--$Y$ plane ($j_z$), and
(d) gas pressure ($p$).
The solid lines depict the magnetic field lines and the arrows represent  
the velocity fields.
Following the initial perturbation,  
the magnetic field lines start to reconnect around the origin.
As the reconnection proceeds, the current
sheet is elongated along the $\pm Y$-direction, 
indicating the formation of a Sweet--Parker current sheet. 
The plasma temperature inside the current sheet $T_\mathrm{cs}$ is
almost uniform with a constant value ($\simeq 8.7 T_\mathrm{in}$). 
If adiabatic heating is the main process for plasma heating,
$T_\mathrm{cs}$ would increase up to $\sim
(\rho_\mathrm{cs}/\rho_\mathrm{in})^{1/3}T_\mathrm{in}\simeq
1.22T_\mathrm{in}$, where $\rho_\mathrm{cs}$ is the density inside
the current sheet. Here we used the numerical result that
$\rho_\mathrm{out}\simeq \rho_\mathrm{cs}\lesssim 1.8\rho_\mathrm{in}$.
Since the evaluated temperature is much less than the observed value,
the plasma should
be heated by ohmic dissipation rather than adiabatic heating
in the diffusion region.
The heated plasma is accelerated in the diffusion region along the $\pm
Y$-direction, resulting in the formation of hot outflows.
The maximum outflow speed at this time is $\sim 0.66$, which is
slightly smaller than the Alfv\'en speed of the sheath plasma ($v_{A,0}
\simeq 0.82$, discussed later). 
The reconnection outflows collide with a magnetic bubble
(plasmoid) that originates in the initial Harris sheet. 
This causes the formation of reverse fast shocks around $Y\simeq 35$ outside
the diffusion region. 
The plasma temperature near the plasmoid increases because of shock
heating and the plasma expands. 
We note that a backflow ($v_y < 0$) structure forms around 
the plasmoid. 
Such backflow structures are also observed in Petschek type
magnetic reconnection, which would be caused by
the expansion of the plasmoid \citep{2010ApJ...716L.214Z}.

At this stage, the electric field around the reconnection region is well
developed.
The typical strength of the electric field near the reconnection point
 ($X\sim Y\sim 0$) is  $E_z \simeq 1.0\times 10^{-2}B_0$.
In addition, 
an electric field ($E_z = v_y B_x$) is also induced near the plasmoid
by the reconnected magnetic fields that are swept 
by the outflow and accumulate around $Y=37$.
Its maximum amplitude is 
$E_z\simeq 0.17B_0$, which is $17$ times larger than 
that near the reconnection point.
Such strong electric fields near plasmoids are also observed in
particle-in-cell simulations \citep{2004ApJ...605L...9J, 2007ApJ...670..702Z} and
two-fluid simulations \citep{2009ApJ...696.1385Z}.
These electric fields can generate high-energy particles 
\citep{2005JGRA..11010215H}. 
\begin{figure}
 \center
 \includegraphics[height=6cm]{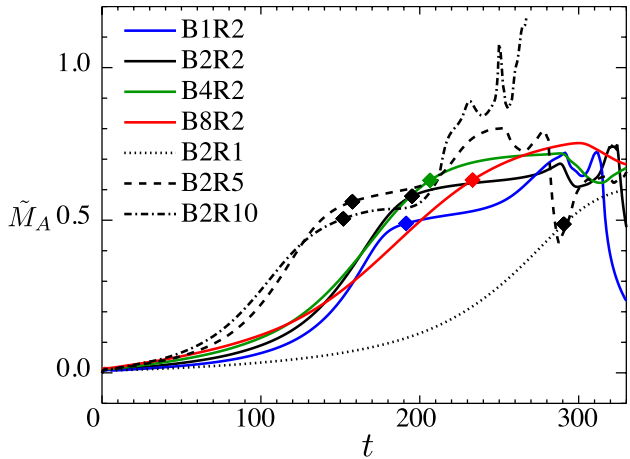}
 \caption{
Time evolution of the maximum Alfv\'en four Mach number
along $X=0$ $({\tilde M}_A)$.
The diamonds denote ${\tilde M}_A$ at $t=t_{20}$.
 }
 \label{fig:fig2}
\end{figure}

Figure~\ref{fig:fig2} shows the time evolution of the maximum outflow
component of the four velocity normalized by the Alfv\'en four speed
in the sheath plasma along $X=0$, $\tilde M_A \equiv
\max[u_y(0,Y)]/u_{A,0}$. 
Each line shows the time evolution of a different model listed in Table~\ref{tab:param}.
After the onset of 
magnetic reconnection, the outflow is accelerated along
the $Y$-direction.
Its acceleration decreases as the reconnection develops (e.g.,
$t\gtrsim 200$ for model B2R2). 
We note that $\tilde M_{A}$ does not increase to
the upper limit of the Alfv\'en Mach number $\tilde M_{A,\mathrm{max}}\equiv
u_\mathrm{max}/u_{A,0}$ listed in
Table~\ref{tab:param} for each model, 
where $u_{\mathrm{max}}=\sqrt{\gamma_\mathrm{max}^2-1}$. 
This suggests that the magnetic energy is not efficiently converted 
into kinetic energy. 
The saturation values of $\tilde M_A$ are independent of $R_M$, 
but weakly decrease as $\sigma_0$ increases. 

In order to estimate the energy composition of inflow and outflow plasmas,
we evaluate the magnetization parameter $\sigma$, specific enthalpy
$h$ and total specific enthalpy $1+\sigma+h$
at $(X,Y)=(5,0)$ for inflows and at $(X,Y)=(0,20)$ for outflows,
summarized in Table~\ref{tab:param}. 
These values are evaluated when the length of the
current sheet $L$, which is defined by $j_z(0,L)/j_z(0,0)=\alpha$,
reaches $L = L_{20}\equiv 20$ (hereafter, we refer to this time
as $t_{20}$). We take $\alpha=0.5$ but the results are independent of $\alpha$
when $\alpha<1$ since $j_z$ is almost constant inside the current sheet and
rapidly decreases at the edge of the current sheet.
For example, the length of the current sheet of model B2R2 at $t=230$ is
$L=33.3$ (see Fig. 1c).

Since we take $\sigma_0>1$, the inflow plasma is dominated by the
magnetic energy ($1, h_\mathrm{in} \ll \sigma_\mathrm{in}$). 
In the outflow plasma, $h_\mathrm{out}$ is much larger than
$\sigma_\mathrm{out}$ and unity, i.e., the thermal energy density 
dominates the rest mass and magnetic energy densities.
Moreover, $h_\mathrm{out}$ is comparable to the total
specific enthalpy of inflow ($1+\sigma_\mathrm{in}+h_\mathrm{in}$).
These results show that almost all the magnetic energy is converted 
into thermal energy due to joule heating. 
By using this fact, we can evaluate the outflow Lorentz factor from equation
(\ref{eq:gout}) as $\gamma_\mathrm{out}\simeq 1$ in the condition
$\gamma_\mathrm{in}\simeq 1$.
The outflow is not accelerated to a relativistic speed
$\gamma_\mathrm{max}$.
This result is consistent with that of the analytical work by
\cite{2005MNRAS.358..113L}. 
They concluded that the magnetic energy should be converted 
into thermal energy ($\sigma_\mathrm{in}\simeq h_\mathrm{out}$) to
maintain the pressure balance across the current sheet.
We confirmed that the current sheet reaches the pressure balance
in the reconnection system, resulting in the formation of relativistically
hot outflows.
Such hot outflows cannot be accelerated to 
a relativistic speed $\gamma_\mathrm{max}$ since the magnetic energy 
is spent in heating rather than acceleration.
Moreover, a larger enthalpy ($h>1$) leads to 
an increase in the plasma inertia in the relativistic regime (see equation
\ref{eq:gout}). 
This also prevents the outflow from being accelerated.
Sub-Alfv\'enic outflow is a natural outcome of relativistic
Sweet--Parker reconnection
because $\gamma_\mathrm{out}=\sqrt{u_\mathrm{out}^2+1}$ remains of order
unity when $u_{A,0}$
increases with $\sigma_0$.  Thus, the Alfv\'en four Mach number
decreases with increasing $\sigma_0$ in the relativistic regime. 
This conclusion is different from that for
non-relativistic reconnection, in which the outflow speed reaches the
Alfv\'en speed.
\begin{figure}
 \center
 \includegraphics[height=6cm]{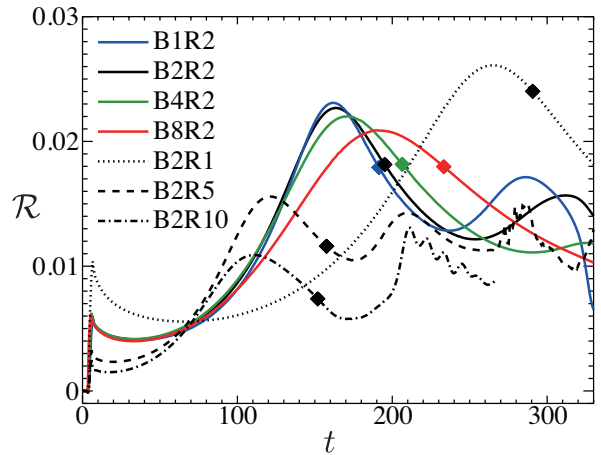}
 \caption{
Time evolution of the reconnection rate ($\mathcal{R}$). 
The diamonds denote $\mathcal{R}$ at $t=t_{20}$.
}
 \label{fig:fig3}
\end{figure}

Figure~\ref{fig:fig3} shows the time evolution of the reconnection rate
$\mathcal{R}=-v_x/v_A$ at $(X,Y)=(5,0)$. 
An initial rapid increase in the reconnection rate is caused by 
the initially imposed perturbation. 
After that, the reconnection rate gradually increases with time and
reaches a maximum. 
Although the outflow speed is almost maintained at the saturated
value, the reconnection rate decreases after passing its peak because
the current sheet continues to elongate. 
Such an elongation of the current sheet is also observed in non-relativistic
plasma \citep{2001ApJ...551..312T}. 
According to the non-relativistic theory, the aspect ratio of the Sweet--Parker
current sheet is described by $\delta /L =\mathcal{R}$. 
Since the curvature radius of the initial magnetic field is infinite
in the Harris sheet (i.e., uniform in the $Y$-direction),
$L$ tends to increase as the reconnection
proceeds, resulting in a decrease in the reconnection rate.

\begin{figure}
 \center
 \includegraphics[height=6cm]{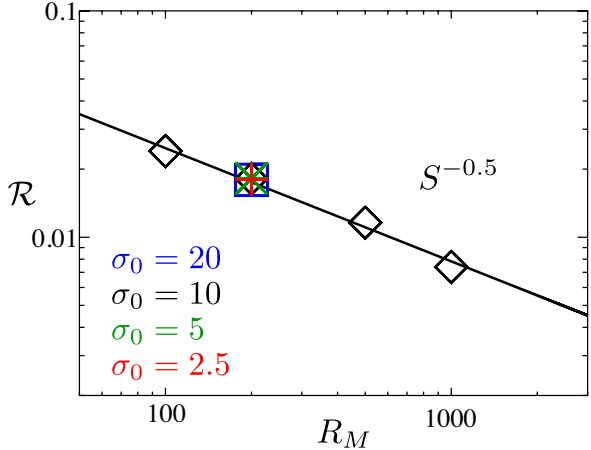}
 \caption{
Reconnection rate $(\mathcal{R})$ as a function of 
magnetic Reynolds number $(R_M)$.
The reconnection rates are evaluated at $(X,Y)=(5,0)$ when $t=t_{20}$. 
The black diamonds correspond to models of different $R_M$ with the same
$\sigma_0 = 10$, while the blue square, green asterisk, and red
cross correspond to models of $\sigma_0=20$, 5, 2.5 with the same
$R_M=200$, respectively.
The solid line shows $\mathcal{R}=S^{-0.5}$, where $S$ is the Lundquist number,
proportional to $R_M$.
}
 \label{fig:fig4}
\end{figure}

Figure~\ref{fig:fig4} shows the reconnection rate $\mathcal{R}$ at
$(X,Y)=(5,0)$ as a function of $R_M$ at $t=t_{20}$. The
reconnection rate is almost independent of $\sigma_0$ for a fixed
$R_M=200$, while it decreases with $R_M$ for a fixed $\sigma_0 = 10$.
The solid line in this figure shows the relation $\mathcal{R}=S^{-1/2}$,
obtained by \cite{2005MNRAS.358..113L} (see equation 8 in
that paper), where $S \equiv 4\pi L_{20} v_{A,0}/\eta =
R_M(L_{20}/\lambda) v_{A,0}$ is the Lundquist number. 
In this plot, we assumed $\sigma_0 = 10$ to evaluate the Lundquist
number that weakly depends on $\sigma_0$.

We found that the reconnection rate is well fitted by the steady
model $\mathcal{R}=S^{-0.5}$, while the system is not exactly steady
state. Although we are not sure why our time-dependent
results follow those of the steady model, we also found the relation 
$t_{L}/t_\mathrm{rec}=0.5$ holds in the Sweet--Parker regime
(e.g., $170\lesssim t \lesssim 240$ for the model B2R2), where
$t_\mathrm{rec}=-(d\log \mathcal{R}/dt)^{-1}$ is the decreasing time
of the reconnection rate and $t_L=(d\log L/dt)^{-1}$
is the increasing time of the length of the current sheet.
This relation is obtained from $\mathcal{R}=S^{-0.5}$ by assuming that
$\mathcal{R}$ and $L$ are time dependent.
This means that the reconnection system evolves
while maintaining the relation $\mathcal{R}=S^{-0.5}$.

\cite{2003ApJ...589..893L} proposed that the
reconnection rate is enhanced by the relativistic effect of the
Lorentz contraction by a factor $\sqrt{\sigma_0}$ when $\sigma_0 \gg 1$
(see equation 39 in that paper). 
In our numerical models, however, such effects never become effective
because of mildly relativistic outflow ($\gamma\simeq
1$), as discussed above. 
Therefore, the reconnection rate is almost independent of $\sigma_0$ and
is described by $S^{-0.5}$.
We conclude that relativistic Sweet--Parker reconnection is a slow
process for energy conversion, as for non-relativistic plasma. 

\section{Conclusion and Discussion} \label{sec:conclusion}
We have developed a relativistic resistive magnetohydrodynamic (RRMHD) code 
that is applicable to plasmas with larger magnetic Reynolds numbers
than possible in previous studies \citep{2006ApJ...647L.123W, 2010ApJ...716L.214Z}. 
We confirmed that the reconnection outflow does not accelerate to 
a relativistic speed
because the magnetic energy released by magnetic reconnection is 
spent on plasma heating rather than acceleration.
The plasma heating results in increasing inertia, 
which also prevents the outflow from being accelerated.
Since the Lorentz factor of the outflow is of order unity,
the enhancement of the reconnection rate due to the Lorentz contraction
is ineffective.
Thus, the reconnection rate of the relativistic plasma obeys the relation 
$\mathcal{R}=S^{-0.5}$, which is the same as that for
non-relativistic plasmas. We confirmed that this relation holds
in the Sweet--Parker regime (i.e., $t_{10}\lesssim t \lesssim t_{30}$,
where $t_{10}$ and $t_{30}$ are the time at which $L=10$ and $L=30$, respectively).
These results are consistent with the
theoretical prediction of \cite{2005MNRAS.358..113L}.

We note that humps appear in the reconnection rate 
beyond the Sweet--Parker regime (see e.g., $t>200$ for the model of B2R10 in
Fig.~\ref{fig:fig3}). At this stage, we observed
a growth of the tearing instability
in the elongated current sheet
(Fig. 1d, the model of B2R10).
An increase in the reconnection rate following the
instability is also observed in non-relativistic reconnection
\citep{2001EP&S...53..473S,
2009PhPl...16k2102B,2009PhPl...16l0702C,
2010PhRvL.105w5002U,2010PhPl...17f2104H}. 
Once the tearing instability develops, the current sheet is disrupted
and the system evolves to a non-steady state. A remarkable feature is
that the increase in the outflow velocity 
coincides with the enhancement of the reconnection rate (e.g., $t>200$ of
model B2R10 in Figures~\ref{fig:fig2} and \ref{fig:fig3}). 
Although the reason for the acceleration is not yet clear, we
speculate that the outflow is accelerated by the pressure gradient force.
When the plasmoid created due to the tearing instability flows along the
current sheet, the plasma in the current sheet is swept up by it.
The subsequent outflow
can then be accelerated by the pressure gradient force (see Fig.~{\ref{fig:fig1}}d). 
Since the plasma inertia $h \sim p/\rho $ decreases while it is
accelerated by the pressure gradient force, 
the outflow speed might be ultra-relativistic $\gamma \gg 1$. 
In such cases, relativistic effects would facilitate the energy
conversion in the reconnection process. These
scenarios need to
be verified through numerical simulations and will be reported in a
subsequent paper.

\acknowledgments
We are grateful to the anonymous referee for improving our manuscript. 
We thank Ken-ichi Nishikawa, Ryoji Matsumoto, Shin-ya Nitta,
Shu-ichiro Inutsuka, Tomoyuki Hanawa and Yosuke Mizuno, for helpful
discussions. 
Part of this work was done while H. R. T. was visiting the University of
Alabama in Huntsville.  Support from the National Science Foundation is
gratefully acknowledged.
Numerical computations were carried out on Cray XT4 at the Center for
Computational Astrophysics, CfCA, at the National Astronomical Observatory
of Japan and on Fujitsu FX-1 at the JAXA Supercomputer System (JSS) at the Japan
Aerospace Exploration Agency (JAXA).

\end{document}